\documentstyle[11pt,newpasp,twoside,epsf,psfig]{article}
\def\arg#1{{\it#1\/}}

\def\edcomment#1{\iffalse\marginpar{\raggedright\sl#1\/}\else\relax\fi}
\reversemarginpar

\begin{document}

\title{Regions of Star Formation:
 Chemical Issues\\ Discussion Session}
\author{{\'A}ngeles I. D{\'\i}az}
\affil{Dpto. de F{\'\i}sica Te{\'o}rica, C-XI, Universidad Aut{\'o}noma de Madrid, Spain}
\author{Mike G. Edmunds}
\affil{Department of Physics and Astronomy, University of Wales, Cardiff, U.K.}
\author{Elena Terlevich}
\affil{INAOE, Tonantzintla, Puebla, M{\'e}xico}

\begin{abstract}
Three are the main questions that were posed to the audience during
this discussion session: a) Can galaxy abundances be believed?, b) What
progress can we expect soon and from where? and c) Can we agree, as a
community, on topics in which effort should be concentrated in the next
five years? 

In what follows, the different contributions by people in
the audience are reflected as they were said trying to convey the
lively spirit that enlightened the discussion. 
\end{abstract}

\section{Introduction}
We proposed to the general audience the discussion of  some of the
following problems: (1) Can R$_{23}$ be trusted as a strong-line abundance 
indicator, particularly for high abundances? (2) Are there other
strong-line type abundance indicators that might be
used at high redshift? (3) How much can infrared nebular diagnostics help?
(4)Should we believe apparent metallicity versus luminosity relations
until the diagnostics are better established?
(5)What are the knock-on effects of the new solar 8.7 oxygen abundance?
(6) What are the REAL accuracies of abundances from X-ray spectra?. Our 
aim was to lead the discussion to these points, incorporating any
others that might arise in relation to them. 

\section{Discussion}

{\bf Mike Edmunds:} First let's talk a bit about the strong line methods for 
analysing HII regions; I think there is a bit
of a crisis at the moment in the calibration of them. Seeing that
people like to study early (i.e. high-redshift)
galaxies nowadays using R$_{23}$, we ought to clear this one up. I often
feel, and Bernard must
feel the same, that the strong line method is 
like a car you have used for a while, and then you sell it on, so it is
a ``used" car and
people keep coming back and saying: look, this car you sold me isn't
working very well, could you 
fix it please?
But no, you bought the car, you fix it yourself! 
For the strong line method, there is a useful paper by Kewley and Dopita (2002)
which gives a sort of DIY kit, lying out the different
line ratios from their models, and how to  use them. BUT they are
essentially on the old calibration scale and shouldn't be used
blindly.


\begin{center}
\begin{figure}
\psfig{figure=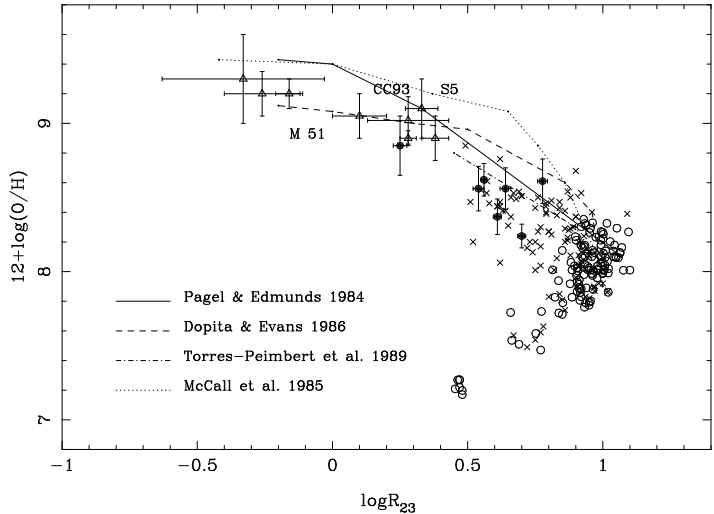,angle=270,width=8cm,rheight=8cm}
\caption{log R$_{23}$ {\it vs} oxygen abundance. Solid dots correspond
 to HII regions with electron temperatures derived from the [SIII]
 $\lambda$ 6312 \AA{} line (Castellanos et al. 2002)}
\end{figure}
\end{center}


\vspace*{-1.15cm}
The problem is as follows: here are some of Angeles' and Elena's and
others data,
the classic diagram of log R$_{23}$ against O abundance (fig. 1), and basically, most
people would agree that the branch at LOW abundances 
is OK. If it is the appropriate branch, and you place strong lines on
it, you will get a reliable abundance out. The
problem comes if you move out to the upper, high abundance branch of
the diagramme, which may OVERESTIMATE the abundance.  Recently,
Pilyugin (2001),
 has used ONLY
HII regions where he can find a 
measurement of the weak temperature-sensitive [OIII] $\lambda$4363 {\AA}  line in
the literature.
These are mainly HII regions in the Galactic disc (Caplan et al. 2000;
Deharveng et al. 2000). The calibration gives R$_{23}$ deduced
abundances that are
considerably lower than the calibrations using HII region models,
although the 4363 calibrators
do not extend very far along the oxygen abundance axis, because of the 
vanishing of the 4363 line in metal rich HII regions.
So the question is, should we revise the scale now using only 4363 calibrators
or should we stick to HII region models?
Angeles has managed to measure an [SIII] temperature in a ``modelled" HII
region (Castellanos, D\'\i az \& Terlevich 2002) and this region
lies above Plyugin's calibration. So the truth may well lie somewhere between
Pilyugin's calibration and (say) the Edmunds and Pagel (1984) calibration - 
and we would concede that the latter must be too high. It is an important
topic, since the differences could be a factor of three at the high 
metallicity end. At least everybody agrees that the systematic trend of
R$_{23}$ with abundance essencially comes from the effective temperature of the
ionizing stars decreasing with increasing metallicity. So why do 
these two calibrations differ? One thing that worries me a little bit
is that with 4363  you are trying to measure something that goes
smaller and smaller and smaller as metallicity increases, and this line
tends to be the one measured
with lowest flux and poorest accuracy in the HII regions. So I just
wonder if there is a bit of a Malmquist bias
in that you measure something that you think is a 4363 line and the
error distribution will always tend to push you to the low
metallicity side - because the region will be dropped from the sample
if 4363 is not seen. What else
could be wrong? Are the Caplan and Deharveng
galactic regions not typical Giant HII regions as you see in the
outside galaxies?
What's wrong with these photoionisation models? People have made them
for years, but are we going to hear  of better
photoionisation models which will drift back down the diagram? Even if
we try and push this with
larger telescopes to better measurements of 4363, Grazyna Stazinska has
recently suggested that 
you may have problems anyway using 4363, it may cause underestimate of the 
true abundance because of temperature gradients. So what is wrong? That
is really what we want to ask. Perhaps Bernard knows the answer.... 

{\bf Bernard Pagel} You compared R$_{23}$ to a used car... I would prefer to give you a
quotation from Macbeth:
``Bloody instructions which being tought, return to plague the inventor".
Now, I am not too worried about the Malmquist bias point, for the
following reason: we calibrated the 
thing originally, on the basis of a model of S5, in M101, by Shields
and Searle, and that S5 has been
measured by Rosa, and  by Kinkel and Rosa, 
and they got a considerably lower O abundance, so, since that region
had already been selected, 
I don't think the Malmquist bias is that important. But I think your
last point is quite correct, 
and there is a lot of work, some of which Manuel alluded to in his
talk, where they now manage
to see recombination lines in these EGHII regions, 30Dor and in
NGC5455, 5461 and NGC604, I believe, 
and they detected recombination lines, and they get a somewhat higher
abundance from the recombination lines
than from the forbidden lines by the conventional methods, so I think
that that last point is a serious one
and it will tend to push the calibration back up again, that's all.

{\bf Angeles D\'\i az:} Well, just looking at this problem at high metallicity, I
think that at the highest 
metallicity end
we will still have a problem when the [OIII] line becomes very weak. In
the intermediate metallicity range
everything works fine, because it is where the O lines are contributing
most to the cooling, but
if we try to measure higher  metallicities, maybe we should turn to 
diagnostics or to lines which
are dominating more the cooling at high metallicities (lower
temperatures). I thought initially
about the [SIII] lines because of that, also because they are less
sensitive to electron temperature,
and I thought that it might solve in some way, or at least help to
solve, the problem. I would also like
to make the point that, although high metallicity is a problem, when 
using the R$_{23}$ relation for HII galaxies, where we think we don't
have a problem in
determining empirical abundances,
most of the objects lie on the ``turn-over" region, and we are (well, I
don't know if
happily) applying this sort of relation to those objects even at
intermediate redshift; so the idea was 
to try to get this ``turn-over" part of the plot into a linear form, and
I think that the S$_{23}$ calibration helps you to do that (see Fig. 2). 

\begin{center}
\begin{figure}
\psfig{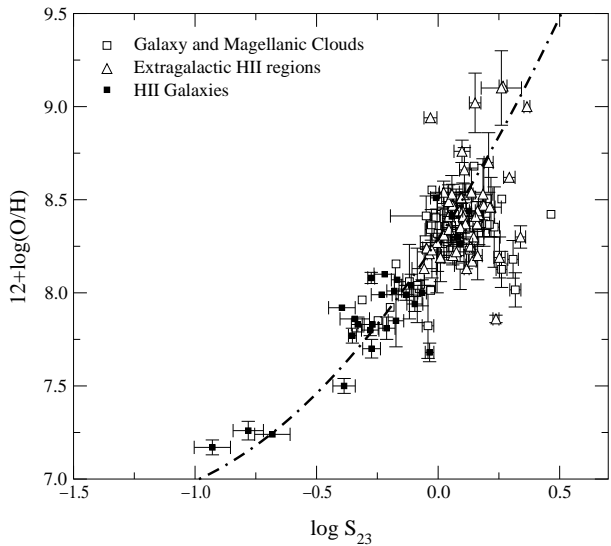}
\caption{log S$_{23}$ {\it vs} oxygen abundance.}
\end{figure}
\end{center}

\vspace*{-1.15cm}
Another reason why I propose to use this 
calibration is that, when you use photoionization models with
different parameters (and one has to 
take into account that the highest the metallicity, the more degenerate
is the problem), there is no one mechanism that is dominating, none of
the three parameters is dominating,  so
stellar effective temperature, ionization parameter and O abundance,
all contribute substantially.
When you work with models, you usually get less dispersion with S$_{23}$ than with
R$_{23}$, maybe because of the
lower sensitivity to the electron temperature of the S[III] lines, and 
perhaps that makes them safe.
 
One of our questions is if there are people here who have any ideas for
other diagnostics, maybe 
lines in the IR, for high Z regions.

{\bf Manuel Peimbert} Getting back to what Bernard was saying, I mentioned in my talk
that there is already data
for about 10 HII regions with recombination lines of O and the point
here is that for the 
10m telescope of the Canary Islands, it will be important to have a 
spectrograph of intermediate to 
high resolution to get these lines, so with a R of 6000 to 9000 I think
it will be possible to have
hundreds of HII regions with these lines detected, and I think this
will improve a lot
the calibration. And of course the calibration is needed to use it in
objects where we cannot
observe these lines. Once it is calibrated, you can go to very high redshifts.

{\bf Don Garnett:} I would also like to push the IR lines as well, because that will
solve a lot of the debate
over whether recombination lines or
collisionally excited lines give the answer we want. The differences
are actually not as large as
they seem.
What we really need is an independent diagnostic, we need [OIII] 88$\mu$
line measurements and even 
the mid IR [SIII] will provide a useful diagnostic as well. The answer is
Optical/IR comparisons
to really push this, to really solve the problem.

{\bf Mike Edmunds:} OK we are moving along. The new solar O abundance 
(Allende Prieto et al. 2001). Everybody was a little
surprised I think when that went
down rather than Orion going up, but there we are. What effect will
that have in the global HII 
region measurements? Bernard, would you like to comment on that?

{\bf Bernard Pagel:} I was actually rather pleased with this result,
because it 
produced consistency between the Sun and both
galactic supergiants and the supergiants in the Magellanic Clouds. 
I've long been maintaining
that the chemical evolution of the Magellanic clouds
could be understood on the basis of the same yields as the chemical
evolution of the local galaxy. So 
actually I was quite pleased at the new O abundance and what you will
also notice is that there is a new C abundance
that brings the C/O ratio back to where it was: 8.4 to 8.5 of O  and
the authors also 
predict that they are going to bring the N down, so the ratio N/O will
be the same. What the rest of the table of abundances shows,
without asking you  to look in detail at it, 
is rather a good consistency between stars and HII regions and SNR in
the various galaxies of the 
Local Group, so the upshot of it is, that the HII region abundances
remain the same, and the Sun
and the stars now finally seem to be catching up with them. But there
are implications because all the differencial
stellar [O/Fe] abundance analysis that has been done
now has to be looked at again, in the light of the new solar O. The obvious
reason for this is
the blend of the [OI] line with a Ni line, which will have different
effect in stars with different metallicity, so that 
complicates life in some respect.

{\bf Rob Kennicutt:} I would like to make a very brief comment, and then a
question to you Bernard.
With Bresolin and Garnett, we are doing an abundance study on M101 
(pre-print). We have electron 
temperatures for 21 HII regions going out from 0.5 to I think 43 Kpc
from the centre and we see this offset of about 0.3
dex between the temperature direct abundance  forbidden lines ($<t^2>$=0 
abundances) and the  Kewley \& Dopita 
calibrations across the board. In fact in your plot you may have
noticed the old Dopita calibration
models are quite different from, say what you have done in the past. My
question to you relating to that
is: the abundances that you have put up, this agreement works if you
use $<t^2>$=0 classical, auroral 
line abundances. If in fact those are wrong, due to fluctuations or
whatever by a factor of 2, won't this
consistency go away?

{\bf Bernard Pagel:} That is not quite true. The Orion abundance was based on 
recombination lines, plus a small
correction for O locked on dust grains (Manuel was one of the authors
with Esteban et al.). Now the 
HII region abundances in M31 and M33 were I think largely based on our old R$_{23}$
method, and so that just leaves 
the LMC, SMC and NGC6822 OK.
Well the implication there is I think that the effective temperature 
fluctuations are
probably not very large. While I'm here I would like to make one other
point: I don't like any attempts 
at calibrating R$_{23}$ or any other line-ratio for abundances, based on a 
sequence of models, partly
because the models have inputs like stellar atmospheres and so on as we
heard this morning,  which are
very complicated things, and you can't always tell what they are going
to do, and partly  because
HII regions don't necessarily form a smooth sequence. For example Evans
and Dopita some time ago assumed a 
sequence in ionization parameter, and McCall, Ribsky and Shields
assumed a sequence in effective 
temperature, and I don't think that the HII regions actually know about
all that. Further more it's
also a question of regression. When there is a dispersion, a regression
of x on y is different from
the regression of y on x. And if you are trying to find an abundance
then it's the regression of x on y 
that you need, not the regression of y on x which people compute from 
sequences of models. And that's 
why I like drawing two straight lines with a corner much better than
drawing continuous curves which
I think can give very misleading results in certain parts of the diagram.

{\bf Elena Terlevich:} So, high redshift, use R$_{23}$? Others? what else would you like?
I thought perhaps we could ask Sara Ellison to tell us something about 
abundances at high redshift. Would you like to? And meanwhile, Gary 
(Steigman), would you like to ask your question? Or suggest your solution? 

{\bf Gary Setigman:} Well, since I'm only an accompanying spouse naive about the
subject, let me ask
a naive question. We have O abundances in stars, but we suspect 
that the abundances on the surface of the Sun are different from the 
abundances that the Sun started with, 
and it's likely that this is true not only for He (which we know from 
helioseismology) but even for elements like O.
So in comparing abundances in stars, don't we have to worry about the 
settling of the heavy elements?
And for the gas, don't we have to worry about how much is tied up in
dust, and whether it's the same 
everywhere in the Universe? I would have thought that the uncertainties
are large enough, that any
result might be obtained, whether you say that the gas and stars agree
or desagree.

{\bf Mike Edmunds:} On the dust one, I think there is some evidence, which I'll talk
a little bit about on
Saturday, that a fairly constant fraction of metals is tied up in dust 
throughout the Universe - I mean must 
vary a bit, but constancy may not be too bad an assumption. 

{\bf Gary Steigman:} Do you know the dust is in the same place as the O lines?

{\bf Mike Edmunds:} No! I'll shut up at this point.

{\bf Bernard Pagel:} I just would like to say something regarding the
settling. People have thought about the 
settling quite a lot. And the settling of the He is OK, it's like about
3 or 4 \% and now you'll notice that 
I only quoted the abundances to the first decimal, because I found
that, you know, quoting the 
abundances to 2 places of decimals is sort of like a second marriage, a
triumph  of hope over
experience. And my feeling is that in the first place of decimals,
the effect of settling in stars like the Sun after convection zones and
so on, is probably not 
something you have to worry about.

{\bf Sara Ellison:}Last night, Elena actually took advantage of my weakenned state, 
having just got off an aeroplane 
after many hours and having had a few sherries, and asked me if I would
say something about high redshift 
abundances. I agreed - that was the effect of the sherry.
My own biased view of high redshift abundances means damped Lyman
alpha systems. These are the heavy-weight 
end of the HI column density distribution of quasar absorption line
systems. Max Pettini, and 
I and many others in many countries, have been looking at different  
abundances in DLA systems. 
Even though most of what we've heard already today has been focused on
low {\em z} systems, there have been
a few DLA abundance results sneaking in there which you might not have 
noticed. For example:
the work that we saw about primordial abundances from Manuel, there
were some D results in there
that have come from Ly limit systems and  DLAs. Also, the work that
Gary was showing about the N/O
abundances  from Max Pettini's work, this also comes from DLAs. There
is a very intimate connection
between what we can do with chemical abundances at low and at high
redshifts. In essence we are trying 
to answer the same questions, about the first stages of chemical
evolution. I could fill up hours talking 
about the abundances in DLAs, but I am trying to keep it short, I don't
want to keep you from your ``copa de vino''. The advantages that we
have when we measure abundances in
DLAs are many-fold. For example this 
whole discussion here  about R$_{23}$ and calibrating the empirical methods,
we don't have to worry about. We can directly determine abundances of
many different elements through direct methods, effectively
measuring equivalent widths or through fitting Voigt profiles. We can
cover  
dozens of different elements using the rest UV resonant
transitions which at high redshift are 
shifted into the optical, and we immediately have the advantage that we
can use ground based
telescopes and detectors that are very efficient, whereas people who
are studying the local ISM are 
struggling in the UV. We don't have to worry very much about ionization
corrections, although how much we 
have to worry about them is still a little bit contentious. One thing
we do have to worry about is dust 
depletion, because we are measuring gas phase abundances, and some
relatively unknown fraction of each 
element is going to be depleted onto dust. So having said all of that,
you can really sum up our
knowledge of DLA abundances in a nutshell by saying that these are
chemically inmature systems. We are seing the first stages of chemical 
evolution. I noticed that
Elena is going to be talking about this ``long and arduous search for
very metal poor galaxies"; the DLAs
are a great test bed for this kind of questions, because they remain
metal poor at all {\em z} that we have 
been able to measure so far. The N/O abundances are low, there is no
evidence for an enrichment of
$\alpha$/Fe peak elements, so all of these are indicators that these are 
chemically inmature galaxies.
I know Elena, Angeles and Mike have circulated a wishlist of what 
theorists and observers
and instrumentalists would like to see coming, or the questions that we
would next like to address.
Really from the point of view of quasar absorption lines, this is a
subject that has really advanced in 
lockstep with instrumentation. The first big step in terms of results
coming from quasar absorption lines 
came with the introduction of the IPCS, when we could first get good
spectra of the quasars and really 
measure these absorption line systems. The next big advance came with
echelle spectrographs on 8m
telescopes, so now with instruments like HIRES on Keck and UVES on the
VLT, we have got  large
abundance databases for around 50 to 60 DLAs. I would say we are
reaching the limit now of what we can 
usefully do at that level, and the next instrumentation step I think it
has to be made, is actually
stepping back from the high resolution point of view. We are using  
resolution of 40-50000 now, so the 
next step I think is going to come from turning down the resolution,
and turning up the wavelength 
coverage. We will still need  the 8-10m class telescopes, so we can go
to much fainter quasars, and we 
can cover many more lines, and the same lines at different redshifts as
well. Keck already has the ESI, and
there is talk about a similar, yet more advanced, instrument for
VLT. I think this is
going to be very important, because at the moment the abundances that
we have are very detailed, but of 
relatively very few systems. Now we are going to increase the number   
that we have in those databases so we can do complete chemical
profiling for many more systems. I also 
think, in touch with what we have heard today about low redshift
systems, making the connection
between the high redshift and the low redshift universe is very
important, because at the high redshift 
we are still ``shooting blind" - we are not so sure what systems these
DLAs represent. We 
know about their chemistry, but it's hard to make the link at the
moment to the low redshift universe.

{\bf Mike Edmunds:} Can I comment? There is a problem with the DLA systems. It's like
the old joke about the drunk man looking under the streetlight for the
keys he lost elsewhere. We have tended to look where we can see, not
where most of the action is happening at high redshift - which  
in chemical terms is in things such as  the forming giant
ellipticals. DLAs are a valuable window into this era, but only a
partial window.

{\bf Sara Ellison:} It's a good point, and that is one of the reasons why I focus on
the point that DLAs are good 
laboratories for metal poor galaxy studies, where you want to ask
questions about the EARLY stages
of chemical evolution. It's true that DLAs are probably not
representative of the global galaxy population, but is a different line
of evidence.
It all depends on what question you want to ask. If you want to ask
questions about 
how the  N/O ratio is varying in chemically very young galaxies, then
DLAs are a good place to look.
If you want to know how global metallicity is evolving from redshift 5
to redshift 0, DLAs are probably 
not the best place to look.

{\bf Alec Boksemberg:} To follow up very slightly on what was just said about DLA
systems, there are other laboratories 
too, which are pretty good and they are the non-damped systems. It is 
possible to do quite a lot
with those and perhaps they don't have some of the limitations, for
example they may not be so selective, 
in the sense that they probably don't have that much dust and they have
a broader range of properties. I think the whole 
combination of absorbers should be looked at, not just selecting
particular ones. Although the DLAs are 
strong objects in attacking the problems, they are not the only ones.

{\bf Brian Boyle:} I'd like to follow-up on Sara's comments about instrumentation
leading a lot of this. Quite right, 
but no one has raised the issue of actually looking at a large sample
of quasars using the
multiobject facilities as are now available in the 8m. The new 
instrumentation coming along will give you
reasonably high resolution over quite  wide fields. The quasar surveys
now in existence (the 2dF quasar survey, the SLOAN digital sky survey) 
are now giving the surface density of potential targets
to allow you to actually get large number statistics in a wide variety
of AGNs, addressing the problem 
that Alec has just raised here. It is not just the DLA systems that are
going to give
you answers, but all absorption systems. So I wonder if Sara could
comment on the use of those sort of 
facilities that are coming on line now, to make the next stride forward.

{\bf Sara Ellison:} It's a very good point, because we do really need to bump up the 
statistics. I mentioned about 
turning down the resolution in order to cover more ``objects" (and/or
lines) but it's a fine balance. Using 
surveys like the 2dF quasar survey and SLOAN is great for pre-selecting
your systems. But obviously you 
cannot study anything like abundances in detail. So I think it's going
to be a very important resource 
for selecting systems for follow-up with higher resolution instruments
which is again going to underline
why we need a moderate resolution instrument, because with these very
large databases you cannot
go through just one per night with UVES.

{\bf Claus Leitherer:} I would like to add another dimension to the discussion of abundances
at high redshift - and that is measurements
directly from the stars, because at high redshift we are in the
fortunate situation that the rest frame UV
is redshifted into the optical. This allows us to define metallicity 
sensitive indices, pretty much
the same as the Lick indices we are all used to, which are mostly
sensitive to metallicity and nothing 
else. Now, from the modelling point of view, this is relatively simple 
because they come from B
stars and there are relatively few uncertainties involved.
Now as you can imagine, in terms of observations, a couple of years ago
we would have said this is
virtually impossible, but now there are some 40 or 50 galaxies
available (I am doing this in collaboration
with a Cambridge group, Samantha Rix and Max Pettini) where we get a
S/N which is sufficient to
measure equivalent widths of the order of an Angstrom (which may sound
like a lot, but in the UV rest 
frame it is not a lot). This allows you to compare to models and you
can determine abundances
to better than a factor of 2, which should be a very useful 
cross-calibration for comparison with the R23 method.

{\bf Elena Terlevich:} We would briefly want to touch on two more points that we have
been thinking about. Are there any  
X-ray astronomers  prepared to admit they are here? Hagai very kindly
has agreed to comment on X-ray abundances.

{\bf Hagai Netzer:} I'm not going to get you to your wine quickly, I'm afraid! We
have heard about
galactic wind superbubbles etc, but I think we heard very little about 
serious attempts to measure
the abundance in the observed X-rays. I want to make a couple of points
here that I think we should all 
bear in mind. It's a very, very difficult job and I must say that NASA,
and perhaps to a lesser
extent, ESA, are doing a great public relation work in advertising
Chandra and XMM-Newton, but most of 
what has been advertised is based on Chandra's superb spatial
resolution and Newton's superb (by X-ray 
standards) spectral resolution. Unfortunately, it's not really very
good for obtaining metallicities
from X-ray images of these winds and bubbles because of the S/N: if we
want a proper S/N to do it, 
unfortunately we have to integrate for many hundreds of thousands of
seconds. One way I like to think about this is that every second 
of Chandra time costs about 50 dollars, so you can work out how
expensive high S/N observations are. So the result is
that what has been published so far (and I want to remind you that
these missions are out there for
three years already, or at least Chandra is) on X-ray metallicities, is
only based on the poor resolution
of these spectrographs, because they are using the CCDs. While they
improved a lot on the imaging 
capability, very little has been done on this. So what I want to push
for, is to say that if these 
instruments are being used properly (in the sense that the community
decides that it's worth spending
a lot of time on) we should spend a lot of time observing a few - and I
mean very very few (it may be 2 or 3 or 4) - of what we all agree, or we 
should all agree, are standard starburst galaxies. Then we must 
see emission lines in the X-rays. Individual emission lines have not
been observed in these objects, or
at least have not been analysed properly.  Needless to say, no one can 
measure the velocity from the
X-rays because the resolution is not good enough. So one thing I'm
almost tempted to say, and let me 
finish with that, is that if everybody in this room is willing to sign
a petition and the petition will be a 
proposal for Chandra to spend a million seconds on one object, I think
that the entire area of star 
formation and these bubbles, and what we can learn from the X-rays,
will benefit a lot. So far, 
I don't think that there is a group of people who were brave enough to 
propose such long exposures,
but this must be done. With present day X-ray instrumentation, the most
one can hope for is to obtain
good abundance analysis on very few number of objects, ONLY if enough
time is spent, and I told you 
how much time is needed. 

{\bf Mike Edmunds:} So don't believe all you read about X-ray abundances accurate to
less than 0.2 dex! It's therefore still well worth
working on trying to refine optical and infrared metallicity indices.

{\bf Angeles D\'\i az:} I will try to end up this discussion by asking all of you: do
we believe things like
the metallicity vs. luminosity relation for HII galaxies and blue
compact galaxies?; do we 
really believe the shape of the abundance gradients which are being
commonly used in chemical
evolution models? Until diagnostics are secure, how reliable are all
these relations we are talking
about?

{\bf Mike Edmunds:} Probably not in detail! But trends may be OK.

{\bf Angeles D\'\i az:} Miguel Mas wants to say one more thing:

{\bf Miguel Mas:} Yes, I want to give a bit more optimistic view of X-ray
spectroscopy than Hagai, in the sense that 
I agree with him about the data obtained with Chandra. Chandra is very
good in its spatial resolution, 
its spectral resolution is very poor. But on Newton we have a
reflection grating spectrometer
and this is another story. The data that has been obtained with this
RGS on Newton for NGC1068, or NGC253
or for M82, are spectacular, are really splendid. These have not been 
published until a few months ago 
and, well, the analysis is really complicated. This is a new tool, a
new technique,  and what we have to do is to learn how to use this new kind
of data which explains why the results have been delayed for
two years. But the results are really spectacular. What we are finding
is that this hot gas is very 
metal rich, the amount of metals is solar or over solar by several
factors. It seems that we are going 
in the right direction. And this has been done with 30 kiloseconds 
integration, which is something
acceptable. This will be flying for ten more years. We will have the 
oportunity to ask for time for 
very long observations: the tools are there.

{\bf Elena Terlevich:} Miguel, spectacular plus or minus what?

\end{document}